\documentclass[twocolumn, showpacs, prb]{revtex4-1}
\usepackage{amsmath}
\usepackage{ amssymb }
\usepackage{subfigure}
\usepackage{graphicx}
\usepackage{float}
\usepackage{color}
\usepackage{comment}
\usepackage{soul}

\begin{document}
\title{Particle-resolved dynamics during multilayer growth of C$_{60}$}
\author{Nicola Kleppmann and Sabine H. L. Klapp}
\affiliation{
Institut f\"ur Theoretische Physik,
Technische Universit\"at Berlin,
Hardenbergstr. 36,
10623 Berlin,
Germany
}
\date{\today}
\begin{abstract}

Using large-scale kinetic Monte-Carlo (KMC) simulations, we investigate the non-equilibrium surface growth of the fullerene C$_{60}$. 
Recently, we have presented a self-consistent set of energy barriers that describes the nucleation and multilayer growth of C$_{60}$ for different temperatures 
and adsorption rates in quantitative agreement with experiments [Bommel \textit{et al.}, Nat. Comm. \textbf{5}, 5388 (2014)]. We found that C$_{60}$ displays lateral diffusion resembling 
colloidal systems, however it has to overcome an atom-like energetic step-edge barrier for interlayer diffusion. Here, we focus on the particle-resolved dynamics, and the interplay between surface morphology and particle dynamics
during growth. Comparing C$_{60}$ growth with an atom-like system, we find significant differences in the evolution of the surface morphology, as well as 
the single-particle dynamics on the growing material landscape. By correlating the mean-squared-displacement of particles with their current neighborhood,
we can identify the influence of the different time scales that compete during growth and can pinpoint the differences between the two systems.

\end{abstract}

\maketitle

\section{Introduction}
Grown structures of nanometre-scale organic molecules are the corner stone of organic semiconductor device constructions. 
The desired morphologies of organic molecules in semiconductor devices range from ultra-thin crystalline films \cite{Proehl2005} 
to islands, nanowires \cite{Geng2008} and crystallites \cite{sugaya:063107}. Devices that require such morphologies include solar cells \cite{Schulze2006, Gunes2007}, and transistors \cite{Park2000}. 
Indeed, it is now well established that the morphology, combined with the type and structure of the substrate \cite{Galiana2013}, determines the device functionality. Their influence
determines features such as the electron transport, the charge carrier mobility \cite{Hlawacek2013}, and the band gap energies \cite{Blumstengel2010} of semiconductor devices.
 These features are, in turn, strongly influenced
by morphological imperfections, which easily arise during the (non-equilibrium) growth process of organic structures on the substrate. It is therefore crucial
to understand on a {\em microscopic} (molecular) level the entire process of formation of such organic-molecule structures from growth towards the final equilibrium state.\\ 
From the experimental side, information on the morphology of the organic component is obtained e.g., via atomic force- \cite{Blumstengel2010, Cao2011} and scanning electron microscopy \cite{Placencia2009}, 
Raman scattering \cite{Khosroabadi2014}, X-ray scattering \cite{Unseres, Hinderhofer2010} and electron microscopy \cite{Heringdorf2001, Khokhar2012}. 
In particular, real-time X-ray scattering can be used to monitor the film formation {\em in situ} and thus gives important information
about the system's behavior on its way {\em towards} thermal equilibrium. 
From the theoretical side, the techniques employed to investigate equilibrium (or even ground state) structures of organic molecules range from ab-initio density functional theory (DFT) 
\cite{Palczynski2014} over atomically resolved Molecular Dynamics (MD) simulations \cite{Potocar2011}  to 
coarse-grained Monte Carlo (MC) simulations \cite{Knychala2012} (see \cite{Clancy2011} for a recent review). Growth processes are typically studied via kinetic Monte Carlo (KMC) 
simulations \cite{Haran2007} and rate equations (see, e.g. \cite{Gibou2003}). \\
The above examples illustrate that the structure formation of organic molecules is a very active field of research.
However, contrary to the situation for atomic systems \cite{Levi1997}, a comprehensive understanding of corresponding organic-molecule systems is still missing.
From the theoretical perspective, one major challenge is the molecule's anisotropy which strongly increases the configurational
space in equilibrium sampling, as well as the space of possible movements during surface growth. Another challenge, particularly for MD and MC simulations relying on classical force fields, arises due to the typically complicated charge distributions and polarizability effects characterizing many organic molecules.\\
In the present paper we investigate, based on particle-resolved KMC simulations, the growth of molecular films composed of C$_{60}$ (fullerene). 
In particular, we aim to explore the {\em single-particle dynamic properties} that accompany multilayer growth. These include free particle diffusion, caging, and detaching from neighbors.
First steps towards unravelling C$_{60}$ multilayer growth have been taken in Ref.~\cite{Unseres} where we developed, together with experiments, 
a KMC model capable of describing various real-space data.\\
Application-wise, C$_{60}$ is a key component to 
semiconductor devices such as transistors \cite{Park2000, Zhang2007} and solar cells \cite{Potscavage2007} because of its high electron yield and photophysical properties \cite{Brabec2001, Arbogast1991}. From a more conceptual perspective, 
C$_{60}$ is clearly one of the easiest representatives within the material class of organic molecules due to its nearly spherical shape. Indeed, at the temperatures considered in our study C$_{60}$
is known to rotate freely not only in the fluid phase, but also in the bulk crystal \cite{Heiney1991} and in one-dimensional confinement \cite{Rols2008}. Thus,
one may expect nearly-free rotations also in film-like geometries.  Moreover, since a C$_{60}$ molecule involves only carbon atoms, partial charge effects are not important.
All these features suggest viewing C$_{60}$ as a particularly large {\em atom} (with a diameter of about one nanometre) rather 
than as a large organic molecule.\\
However, besides size and internal structure there is another important difference between C$_{60}$ molecules and atoms: The range of the {\em effective}, i.e. angle-averaged, attractive interaction
between two C$_{60}$ molecules is much smaller than the usual van-der Waals interaction (decaying as $r^{-6}$, $r$ being the separation) between atoms 
\cite{Tewari2010, Girifalco1992}. This difference between the pair potentials is illustrated in Fig.~ \ref{fig:system_sketch}. The short range of attraction between C$_{60}$ molecules has important consequences 
for the overall equilibrium phase behavior; in particular, 
 C$_{60}$ lacks a liquid phase \cite{Hagen1993} but tends to form a gel phase \cite{Royall2011}.
In that sense,  C$_{60}$ rather behaves like a system of colloids, where short-range attraction (stemming from depletion effects) is in fact quite common. \\
Here we are interested in C$_{60}$ growth. Earlier theoretical studies in this area have focused on aspects such as 
determination of step-edge barriers and potential
landscapes (yielding diffusion rates) from DFT, see, e.g., \cite{Goose2010, Gravil1996}. 
Studies addressing the surface morphology have often been restricted to a coverage of less than one monolayer \cite{Korner2011,Liu2008}. 
Note that this contrasts with the situation for atomic systems where 
growth phenomena for both, monolayers and multilayers have been studied intensely 
for a wide range of systems \cite{Fu2008,Evans20061, Reuter2006}. These studies include even subtle phenomena such as concerted gliding of islands \cite{Karim2006} or 
direction-resolved step-edge diffusion \cite{Voter2007,Teichert1994}.\\
In our recent study we have obtained, together with real-time experiments \cite{Unseres}, a consistent set of energy-barrier parameters for KMC simulations which describe
measurable morphological quantities such as island density and layer coverage as functions of time. Interestingly, these energy parameters reflect again
the intermediate role of C$_{60}$ between atoms and colloids: While the step-edge diffusion barrier is close to what one expects for atoms, the binding energy stemming from attractive
interactions is much smaller, reflecting indirectly the much shorter range of attraction. In the present study, we focus on the similarities and differences between the 
growth of the spherical molecule C$_{60}$ and comparable atomic systems. Using these two system types, we study the interplay between morphology and the single-particle dynamics during growth.
We find significant differences between the two system types. These differences concern the evolution of their surface morphologies, the long-time scaling behavior of the morphology, and the particle-resolved dynamics. 
The differences are traced back to the different time scales competing in the surface evolution, by correlating 
local surface structures with single-particle dynamics.  \\
The remainder of the paper is organized as follows. In Sec.~\ref{Sec:2Model} we introduce the simulation techniques and target quantities considered in this study.
We also propose a way to compare C$_{60}$ and atom-like systems in terms of energy-barrier arguments.
Section~\ref{Sec:3.1Trajectories} gives an overview of KMC results for the surface morphology, followed by a detailed discussion of global (Sec.~\ref{Sec:3.3Global}) and 
single-particle (Sec.~\ref{Sec:3.2Local}) dynamic quantities. The paper closes with a summary in Sec.~\ref{sec:conclusion}.\\
\section{Method}\label{Sec:2Model}
\subsection{Simulation method}
\begin{figure}%
      \includegraphics[scale=1.0]{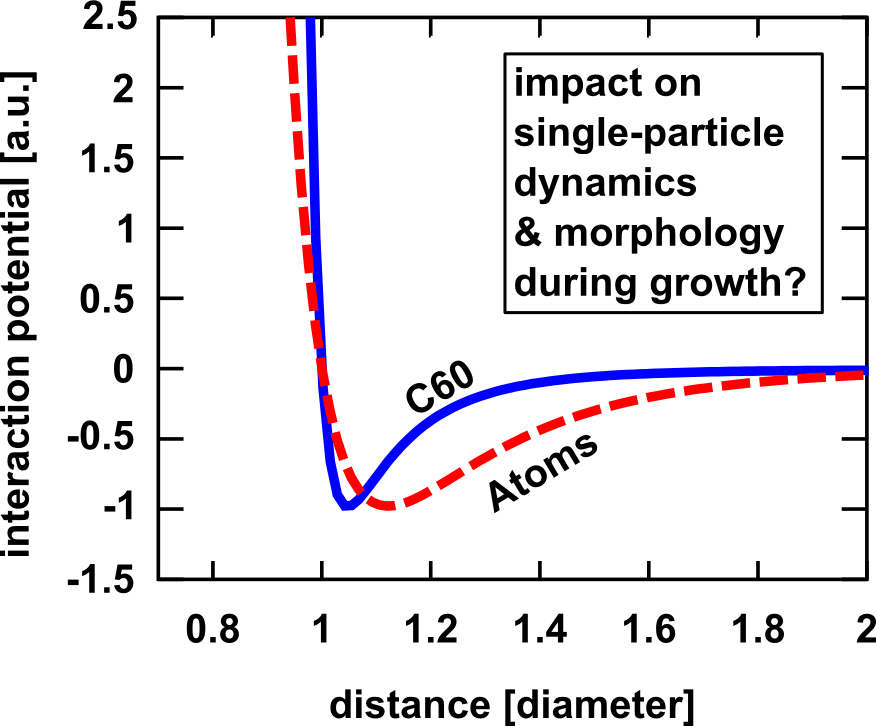}
\caption{(color online) Pair potentials for C$_{60}$ (blue continuous) and argon-atoms (red dashed). 
Both are scaled to a potential well depth of -1, and the range is expressed in terms of particle diameters \cite{Tewari2010, Girifalco1992}.
}\label{fig:system_sketch}%
\end{figure}
We employ event-driven kinetic Monte Carlo simulations, using the N-fold algorithm \cite{Bortz1975} 
(see Ref. \cite{Voter2007} for a review). This allows us to access large surfaces (with $\mathcal{O}(10^6)$ 
lattice sites) with molecular resolution. 
The free diffusion time, that is the average time span a particle takes for one diffusion step if it is not interacting with lateral neighbors ,
is $\mathcal{O}(1\mu\textrm{s})$ (see e.g. \cite{Liu2008}). In total, however, the simulations cover the experimentally relevant range of minutes to hours. \\
During the growth process particles adsorb on the surface
with a constant effective adsorption rate $f=f_{\textrm{adsorb}}-f_{\textrm{desorb}}$ and diffuse on the surface, until they become immobilized due to interaction effects. 
Finally they are buried under the next grown layer. The diffusion process of 
a particle from site $i$ to site $j$ is given through the rate determined in the Clarke-Vvedensky bond-counting Ansatz \cite{Clarke1988, Oliveira2013}:
\begin{equation}	
r_{i,j}=\frac{2 k_{\textrm{B}} T}{h} \exp\left(-\frac{E_{\textrm{free}}+ n_i E_{\mathrm{n}}+s_{i,j} E_{\textrm{ES}}}{k_{\textrm{B}} T}\right)\textrm{,}
\label{eqn::rate} \end{equation}  
where $k_{\textrm{B}}$ is the Boltzmann constant, $T$ is the temperature and $h$ is the Planck constant.
We choose the prefactor $\nu_0=2 k_{\textrm{B}} T/h$ in accordance with previous studies \cite{Gyure1998, Jones2009, Marmorkos1992}. As seen from 
Eq.~(\ref{eqn::rate}), the diffusion rate $r_{i,j}$
depends exponentially on the total energy barrier the particle has to overcome to reach site $j$.	
This energy barrier consists of a contribution for free diffusion, $E_{\textrm{free}}$, a contribution arising from the interaction with the $n_i$ nearest neighbor particles, $E_{\mathrm{n}}$, 
and an Ehrlich-Schwoebel
energy barrier, $E_{\textrm{ES}}$. The latter is relevant for trajectories that lead across step edges, i.e., $s_{i,j}=1$ when crossing step edges, otherwise $s_{i,j}=0$. We note that
the restriction of the interaction term to the nearest neighbors alone is consistent with the fact that the distance-dependent attraction between C$_{60}$ 
molecules is relatively short-ranged compared to the molecular diameter \cite{doi:10.1021/j100167a002}.\\
The simulated time progresses by a time step $\tau$ after each event. This time step is of stochastic nature, and it is weighted with the rate of change of the 
whole system $r_{\textrm{system}}$.
The latter is defined as the sum of all possible process rates $r_{i,j}$ plus the adsorption rate $f$. Specifically, we have
\begin{equation}
  \tau  = \frac{-\ln(R)}{r_{\textrm{system}}}\textrm{, } 
\end{equation}
where $R \in \left[0,1\right]$ is a random number, and
\begin{equation}
  r_{\textrm{system}}=\sum_{i=1}^{N} \left(\sum_{j=1}^{6} r_{i,j}+f\right)\textrm{.}\label{eqn::systemrate}
\end{equation}
In Eq.~(\ref{eqn::systemrate}), $N$ is the number of surface sites in the system. \\
In accordance with experiments \cite{Unseres}, we simulate the growth process on a triangular lattice which is equivalent to the fcc(111) lattice face of a bulk C$_{60}$ crystal. 
Interstitial sites are not considered. Furthermore, we assume that the growth process is free of defects in the sense that particles can only sit on lattice sites and
cannot form overhangs. If particles reach sites that are not supported by three ``base particles'' in the underlying layer, they relax to surrounding lattice sites with 
probabilities proportional to the corresponding diffusion rates $r_{i,j}$.\\
We also note that our simulation does not take into account coordinated, simultaneous motion of particle clusters. Previous studies \cite{Nandipati2009} have indicated
that concerted cluster diffusion influences the growth only during the very initial phase of island nucleation, when the particle clusters are small. 
Furthermore, in a self-learning KMC study of the system Cu/Cu(111), Karim \textit{et al.} \cite{Karim2006} found that the diffusion barriers of clusters scales nearly 
linearly with the cluster size: Specifically, they find an effective diffusion barrier for dimer diffusion, which is
approximately twice as high as that corresponding to monomers. Transferring this trend to our C$_{60}$ system, in which the monomer barrier $E_{\textrm{free}}$ is 
already quite large, 
we expect a very large energy barrier for dimer diffusion. 
Thus, we expect cluster diffusion to take place only on very long time scales and, therefore, we expect very little
influence of concerted cluster diffusion on the dynamic properties studied here. 
The simulation parameters are chosen in accordance with recent experiments \cite{Unseres}, in which \textit{in situ} measurements were made during growth. 
 These experiments use x-ray scattering to gain insight into the real-time evolution of both the island density and layer coverage, simultaneously. The layer
coverage is monitored through the modulation of the scattering intensity at the so-called ``anti-Bragg point'', while the island density is deduced from small-angle
x-ray scattering \cite{Unseres}. By 
comparing these experimentally obtained quantities with corresponding KMC results for a range of temperatures and adsorption rates, 
we were able to find a consistent set of energy barriers. These are $E_{\mathrm{free}}=0.54\pm 0.04\,\textrm{eV}$, $E_{\mathrm{n}}=0.13\,\pm 0.02\,\textrm{eV}$ and 
$E_{\mathrm{ES}}=0.11\,\pm 0.02\,\textrm{eV}$. These parameters have been shown to describe C$_{60}$ for the temperature range $40^{\circ}\textrm{C}-80^{\circ}\textrm{C}$ and adsorption 
rates in the range $\left(0.1\,\textrm{ML~min}^{-1}-1\,\textrm{ML~min}^{-1}\right)$, where ML stands for monolayer \cite{Unseres}. 
We also note that our values are in agreement with previous simulations of a monolayer of C$_{60}$ on C$_{60}$ \cite{Korner2011,Liu2008} if 
modeling differences (concerning our coarse-graining of the lattice, as well as differences
of the definitions of $E_{\mathrm{n}}$ and $\nu$) are taken into account. For a more detailed discussion see Appendix \ref{app1}. In the present study we focus on the temperature
$T=40^{\circ}\textrm{C}$.\\
\begin{figure}%
      \includegraphics[scale=0.45]{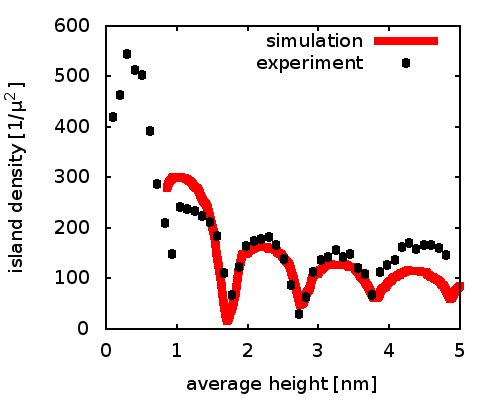}
\caption{(color online) Island density for $T=40^{\circ}$C and a relatively small adsorption rate ($f=0.1$~ML~min$^{-1}$) as a function of the average surface height $\bar{h}=f\times t$
.}\label{fig:island_density}%
\end{figure}
To demonstrate the quantitative agreement between experiment and KMC simulation, we show in Fig.~\ref{fig:island_density} experimental and simulated data for 
the island density as a function of time.
Note that, in the experiment, the first layer of C$_{60}$ was grown on mica, while our simulation begins on the first closed layer of C$_{60}$. Whenever we refer to the first (second, etc.) layer
from this point onward we mean the first layer of C$_{60}$ on C$_{60}$, which is equivalent to the second experimental layer.
\subsection{Target quantities}\label{Sec:2Eval}
The focal point of our study is to explore the interplay between the time-dependent surface morphology on the one hand, and the particle-resolved dynamics on the other hand.\\
To characterize the surface morphology, we calculate the height-height correlation function $G(d,t)$ defined as
\begin{equation}
 G(d,t)=\left\langle\frac{\sum_{i}^{M}\sum_{j}^{M}(h({\bf x}_i,t)-\bar{h})(h({\bf x}_j,t)-\bar{h})}{N(d)}\right\rangle\textrm{,}\label{eqn::hh_correl}
\end{equation}
where $M$ is the number of points on the surface, and $d=\left|{\bf x}_i-{\bf x}_j\right|$ is the distance between two points on the surface, $i$ and $j$.  
These points are characterized by their position vectors ${\bf x}_{i}$, ${\bf x}_{j}$ and their heights $h({\bf x}_{i})$, $h({\bf x}_{j})$. 
The function $G(d,t)$ is determined by averaging over the $N(d)$ pairs of points on the surface, which have a distance $d$, followed by an average over realizations (denoted
 as $\langle \dots \rangle$). 
In Eq.~(\ref{eqn::hh_correl}), $\bar{h}$ is the average height of the surface. The height-height correlation function $G(d,t)$ has successfully been used to characterize a variety of systems, both in experimental studies 
(e.g., in STM-imaging \cite{Karmakar2004L101}) and in simulations \cite{Yu2002}.
The definition~(\ref{eqn::hh_correl}) implies two particularly interesting special cases regarding the values of the distance $d$. The first one corresponds to $d=0$ 
(and thus $i=j$). In this case, $G(0,t)$ can be interpreted as the variance,
\begin{equation}
 G(0,t)=\left\langle\frac{\sum_{i}(h({\bf x}_i,t)-\bar{h})(h({\bf x}_i,t)-\bar{h})}{N(0)} \right\rangle\textrm{.}\label{eqn::hh_correl_0}
\end{equation}
Clearly, the variance is sensitive to deviations from the average surface height $\bar{h}$, which is why it is commonly interpreted as the roughness.
The second special case is that the points ${i,j}$ are nearest neighbors ($i,j \in n$, which is equivalent to $
\left|{\bf x}_i-{\bf x}_j\right|=a$).
Then, we find a measure for the mean squared step height, 
\begin{equation}
 G(1,t)=\left\langle\frac{\sum_{i,j \in n}(h({\bf x}_i,t)-\bar{h})(h({\bf x}_j,t)-\bar{h})}{N(1)} \right\rangle\textrm{,}\label{eqn::hh_correl_1}
\end{equation}
which correlates the heights of neighboring sites. Therefore, $ G(1,t)$ is the correlation function that is most sensitive to local variations; hence it is often 
called the local roughness or mean square step height \cite{dasSarma1996}.\\
A characteristic feature of the present growth process is the formation of islands. As soon as these are present, the surface can also be characterized 
through scalar morphological descriptors such as the fractal dimension $D$ \cite{Schaefer1984}. 
The latter is determined through the scaling behavior of island surface $A$ with the radius of gyration $x_{\textrm{gyr}}$
\begin{equation}
 A\propto x_{\textrm{gyr}}^D \textrm{,} \label{eqn::def_D}
\end{equation}
where $x_{\textrm{gyr}}^D$ is defined as
\begin{equation}
 x_{\textrm{gyr}}=\sqrt{\frac{1}{N^{\textrm{island}}} \sum_{i \in \textrm{island}} \left( {\bf x}_i-\bar{{\bf x}}_{\textrm{island}}\right)^2 }\textrm{.} \label{eqn::r_gyration}
\end{equation}
In Eq.~(\ref{eqn::r_gyration}), $N^{\textrm{island}}$ is the number of particles in the island, and $\bar{{\bf x}}_{\textrm{island}}$ is the center-of-mass position of the island, i.e.
\begin{equation}
 \bar{{\bf x}}_{\textrm{island}}=\frac{1}{N^{\textrm{island}}}\sum_{i \in \textrm{island}} {\bf x}_i\textrm{.}
\end{equation}
The island is defined using a cluster algorithm (Hoshen-Kopelman algorithm, see \cite{Hoshen1976}) to identify all particles within 
one island.\\
The fractal dimension $D$ describes how branched structures are: The closer to two the fractal dimension of a 2D island is, 
the less dendritic is its morphology. Because the island size increases in discrete steps, the scaling behavior of $a$ [see Eq.~(\ref{eqn::def_D})] breaks down for small islands.\\
So far we have focused on system-averaged quantities. To understand the dynamics on a particle level, we analyze the mean-squared displacement (MSD) 
of particles as a function of time. As the particle dynamics depend on the morphology of the substrate, only particles that arrive 
on the surface at time $t$ after beginning of surface growth are considered in the average over realizations (e.g. after the growth of 0.5ML). We then define the MSD as
\begin{equation}
 \Delta x(t^{\ast})^2=\langle\left| {\bf x}(t^{\ast})-{\bf x}(0)\right|^2 \rangle_{t}\textrm{,}
\end{equation}
where ${\bf x}(t^{\ast})$ is the position of the particle at the time $t^{\ast}$ after it's arrival on the substrate, and $\langle ... \rangle_{t}$ is 
the average over all realizations for particles that arrive at time $t$. For free diffusion the MSD scales 
linearly in time, while for immobile particles it assumes a constant value. 
If the MSD scales slower (faster) than linearly with $t^{\ast}$ is called sub-diffusive (super-diffusive) \cite{Metzler2000}.\\
To interpret our MSD results, we look at the processes occurring at time $t^{\ast}$. 
 The process types considered are 
free diffusion, diffusion away from sites with neighboring particles, diffusion across step edges and also immobilization. Immobilized particles are embedded and 
remain immobilized for the rest of the simulation. We define $N(p,t^{\ast})$ as the fraction of particles that perform a specific process of type $p$ at time $t^{\ast}$.
%
\subsection{Systems under investigation} \label{Sec:atomic}
\begin{table}
\begin{tabular}{c|ccc|c}
 \hline \hline
  System & $E_{\textrm{free}}$ & $E_{\textrm{n}}$ & $E_{\textrm{ES}}$ &  $E_{\textrm{n}}^{,}$\\
  \hline
  C$_{60}$ & $0.54\,$eV & $0.13\,$eV & $0.11$\,eV & $0.13$\,eV \\
  Ag & $0.067-0.12$\,eV & $0.19$\,eV & $0.28-0.3$\,eV & $0.72$\,eV \\
  Pt & $0.26$\,eV & $0.5$\,eV & $0.08$\,eV & $0.92$\,eV \\
 \hline \hline
   \end{tabular}
 \caption{Energy parameters used for the KMC simulations of C$_{60}$ \cite{Unseres} and two atomic systems Ag (Refs.~\cite{Evans20061, Nandipati2009,Li2009, Brune1995, Latz2012, Li2008, Blackwell2012}) and Pt (Refs.~\cite{Evans20061, Hohage1996, Feibelman2001L723}). 
 The interaction energies  $E_{\textrm{n}}^{,}$ are discussed in Sec.~\ref{Sec:atomic} as well as in Appendix~\ref{app2}.}
 \label{tab:1values}
\end{table}
A focal point of our study is to contrast the growth dynamics of C$_{60}$ against representative atomic systems. 
For the latter, we choose Pt grown on Pt(111) and Ag grown on Ag(111).
For C$_{60}$, the energetic barrier stemming from the interactions with the nearest neighbors is relatively small; an effect which we explain through the fact that the attractive center-of-mass
interactions between two C$_{60}$ molecules have a rather short range (as compared to atomic systems). The atomic and C$_{60}$ pair-potentials are depicted in  
Fig.~\ref{fig:system_sketch}. The potentials are scaled with the particle diameter and the potential well depth in order to visualize the difference in range. 
As a consequence, the ratio between the energy barrier for 
in-plane diffusion, on the one hand, and the total energy barrier for a particle to break from a dimer, $E_{\textrm{n}}+E_{\textrm{free}}$, on the other hand, is relatively large. Specifically,
we find (see table \ref{tab:1values})
\begin{equation}
 R(\textrm{C}_{60})=\frac{E_{\textrm{free}}}{E_{\textrm{free}}+E_{\textrm{n}}}\approx 0.8 \textrm{.} \label{eqn::R}
\end{equation}
 This large value of $R$ is the major effect of the size of C$_{60}$ and its small interaction range on the energy parameters 
$E_{\textrm{free}}$, $E_{\textrm{n}}$ and $E_{\textrm{ES}}$ \cite{Unseres}. 
We intend to isolate, within our KMC simulations, the role of neighbor interactions on the growth of representative atomic systems relative to C$_{60}$ growth. Thus, 
we proceed as follows: The atom-like KMC simulations are performed with the same values of $E_{\textrm{free}}$, $E_{\textrm{ES}}$ used in the C$_{60}$ simulations. We also assume the same 
lattice configuration and experimental input parameters for the atom-like simulations.
However, the values for the neighbor interaction $E_{\textrm{n}}^,$ of atomic systems are chosen such that the ratio 
$R=E_{\textrm{free}}/(E_{\textrm{free}}+E_{\textrm{n}}^{,})$ fulfills the literature values of $R\approx0.37$ for Pt and $R\approx 0.43$ for Ag. The values used in determining
these ratios are listed in Table~\ref{tab:1values} and discussed in Appendix~\ref{app2}.\\
Analyzing systems that are identical in all parameters except the ratio $R$ allows us a direct comparison of single-particle dynamics despite the smaller time- and 
length scales of growth in atomic systems relative to C$_{60}$. 
\section{Results}\label{Sec:3Results}
\subsection{Morphology and trajectories}\label{Sec:3.1Trajectories}
\begin{figure}%
\begin{minipage}{8cm}%
      \includegraphics[scale=0.45]{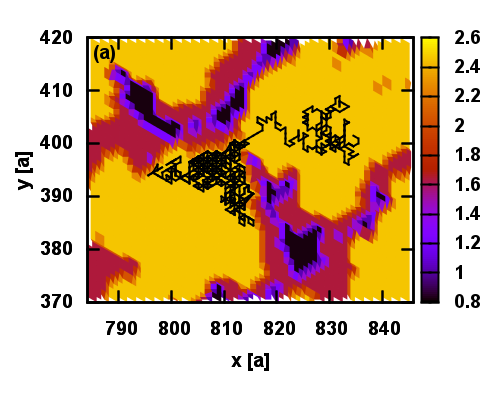}
\end{minipage}\\
\begin{minipage}{8cm}%
      \includegraphics[scale=0.45]{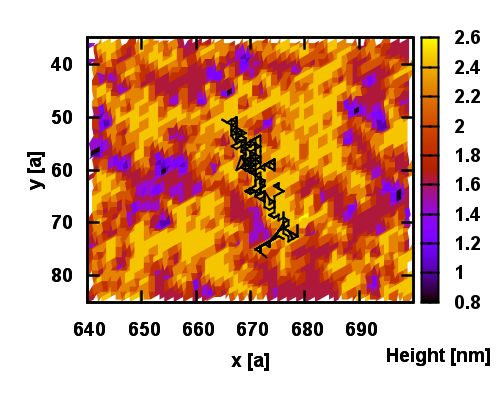}
\end{minipage}%
\caption{(color online) Height of surface structures after the growth of 1.5\,ML for (a) C$_{60}$ on C$_{60}$(111) and (b) Ag on Ag(111). The black lines
depict trajectories of a particle which arrived on an island. Both systems are simulated on a triangular lattice with equal energy barriers 
$E_{\textrm{free}}=0.54$\,eV and $E_{\textrm{ES}}=0.11$\,eV, but different neighbor binding energies $E_{\textrm{n}}=0.13 $\,eV and 
$E_{\textrm{n}}(\textrm{Ag})^{,}=0.72$\,eV, respectively.}%
\label{fig:1trajectories}%
\end{figure}
To start with, we show in Fig.~\ref{fig:1trajectories} two surface structures illustrating the morphology of C$_{60}$ and the Ag system after the growth of 1.5 ML. The adsorption rate here and in the following
figures is $f=0.1$~ML~min$^{-1}$.
All lengths are plotted in units of the lattice constant, $a$.
It is seen that the depicted surfaces have distinctly different structures. Most prominently, C$_{60}$ has well rounded islands while Ag forms dendritic,
nearly fractal structures. This morphological difference reflects the fact that C$_{60}$ has a noticeably higher ratio $R$ than Ag [see Eq.~(\ref{eqn::R})]; therefore, processes 
that break bonds to lateral neighbors are far more likely. As a consequence, particles can easily move to sites with high coordination numbers, which then again 
results in rounded islands. Ag is characterized by a much smaller ratio $R$. Therefore, once particles are bound to their neighbors these bonds are 
less likely to break.
As a result one observes the formation of dendritic structures.\\
In Figs.~\ref{fig:1trajectories} (a) and (b) we have included a typical single-particle trajectory illustrating the individual motion of that particle
after arrival on an island. For both systems, we clearly see that motion across a step edge is hindered.
This leaves the particles to meander mainly on the 
island surface, caged by the island edges. However, as Figs.~\ref{fig:1trajectories}(a) and (b) clearly show, the different shapes of the islands 
influence the shape of the paths. In particular, the rounded islands of C$_{60}$ lead to 
caging into a relatively small surface area within which the particles can diffuse essentially freely. On the other hand, the more fractal structure of the islands formed by 
Ag allows for longer paths of free diffusion (``stretches''), that is, particles move along ``channels'' formed by island edges.
\subsection{Correlation functions}\label{Sec:3.3Global}

In the previous paragraph we have seen that the single-particle dynamics depends crucially on the morphology of the surface. Thus it 
is important to understand the evolution of the surface morphology with time. To this end, we now discuss the behavior of the spatio-temporal correlation functions introduced 
in Eqs.~(\ref{eqn::hh_correl})-(\ref{eqn::hh_correl_1}).\\
\begin{figure}%
      \includegraphics[scale=0.425]{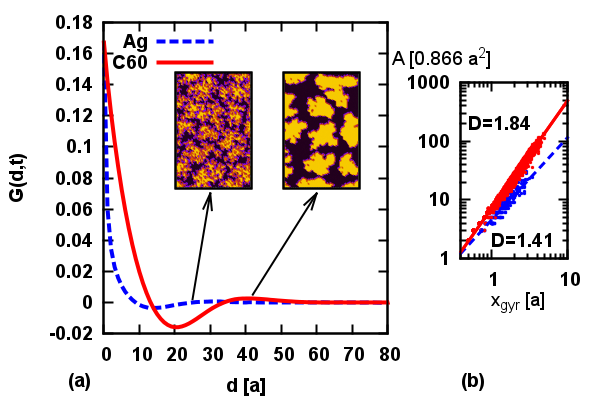}
\caption{(color online) (a) Height-height correlation function $G(d,t)$ as a function of the distance $d$ after the growth of 0.5\,ML ($t\approx 22\,$s) for C$_{60}$ and Ag. The inset
depicts a snapshot of size 75\,$a$ x 125\,$a$ of the surfaces. Part (b) shows the dependence of the average island size on the radius of gyration, $x_{\textrm{gyr}}$, in a double-logarithmic
representation. Included are values for the resulting scaling exponents $D$ [see Eq.~(\ref{eqn::def_D})].}\label{fig:4hh_corr}%
\end{figure}
In the central graph, Fig.~\ref{fig:4hh_corr}(a) depicts the height-height correlation function $G(d,t)$, which correlates the deviation from the average height $\bar{h}$ at 
two points with distance $d$ at time $t$ [see Eq.~(\ref{eqn::hh_correl})]. Specifically, we focus on a time during the growth of the first monolayer ($t\approx 22\,$s). For the Ag system, $G(d,t)$ decays rapidly
to zero. The corresponding function for C$_{60}$ reflects correlations ranging over much larger distances. These differences can be understood using the surface 
snapshots shown in the inset of Fig.~\ref{fig:4hh_corr}(a). The C$_{60}$ system displays clearly separated islands with well defined radii and distances. These features are 
mirrored by strong and long-ranged spatial correlations in the corresponding $G(d,t)$. In particular, the maximum at 42$\,a$ corresponds to the average distance between neighboring 
islands. The Ag system, however, is characterized by a far more dendritic island structure, which is reflected by the short range and smooth structure of $G(d,t)$.\\
The different island structures of C$_{60}$ and Ag can be quantified through the fractal dimension $D$. 
We have determined this quantity via a scaling plot of the island area as function of the radius of gyration (see Fig.~\ref{fig:4hh_corr}(b) and Eq.~(\ref{eqn::def_D}), respectively). From this 
we find $D \approx 1.84\pm 0.01$ for C$_{60}$ and $D \approx 1.41\pm 0.03$ for Ag. The much smaller values for Ag indicates the dendritic morphology of Ag islands.
This confirms our interpretation that the small value of $D^{\textrm{Ag}}$ is due to its strongly dendritic growth. Regarding the fractal 
dimension for C$_{60}$, we note that despite the rather large, ``colloid-like'' value of the energy ratio $R$, our value of $D$ deviates from the fractal dimension characteristically found for (uncharged) colloidal systems 
\cite{Aubert1986, Liu1990}. Rather it lies within the range of values expected for atomic systems.\\
\begin{figure}%
      \includegraphics[scale=0.5]{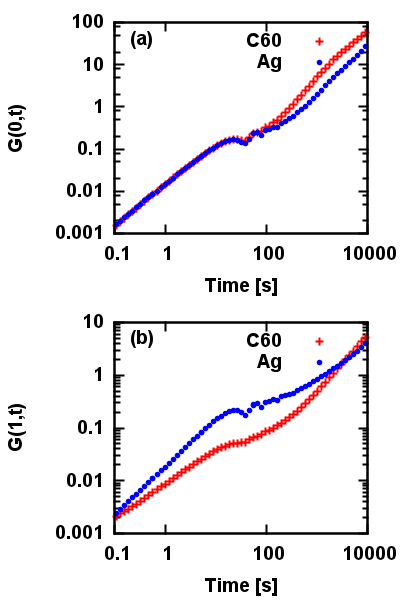}
\caption{(color online) Height-height correlation functions $G(d,t)$ of C$_{60}$ and Ag as a function of time for two distances (a) $d=0$\,$a$, (b) $d=1$\,$a$.}%
\label{fig:5roughness}%
\end{figure}
To further characterize the growth, we show in Fig.~\ref{fig:5roughness} the correlation functions $G(0,t)$ and $G(1,t)$ as a function of time. 
As mentioned in Sec.~\ref{Sec:2Eval}, the functions $G(0,t)$ and $G(1,t)$ measure the degree of overall and local roughness, respectively. 
In systems characterized by layer-by-layer growth the overall roughness initially grows. After $\approx 20s$, the overall roughness saturates at approximately 
$0.2$ and oscillates around this value \cite{Smilauer1995}. At later times, when layer-by-layer growth breaks down, roughening of the surface starts and 
$G(0,t)$ increases again \cite{krug1997}. Indeed, it can increase even indefinitely if mounds or crystallites form on the surface (similar roughness evolution was found, e.g., 
by Smilauer and Vvedensky \cite{Smilauer1995}). In Fig.~\ref{fig:5roughness}(a) we see
that, for the growth of the first layers ($t\lesssim 200\,$s), C$_{60}$ and Ag grow with an identical roughness $G(0,t)$, 
indicating that the systems do not differ significantly in their inter-layer diffusion behavior
and follow similar growth modes. In the subsequent time range $200\,$s$ \lesssim t\lesssim 400\,$s, Ag maintains a constant roughness with a value of about 0.2, while the roughness
of C$_{60}$ progressively increases. We interpret the behavior of Ag as prolongation of layer-by-layer growth, which is characterized by approximately constant deviations of 
local from average surface height. Since the Ag system is simulated with the same Ehrlich-Schwoebel barrier as the C$_{60}$ system, the observed deviation in the temporal behavior of the 
roughness indicates a complex coupling of particle trajectories and 
surface morphology. Finally, for long times ($t\gtrsim 1000\,$s), the roughness of Ag and C$_{60}$ increases with the same exponent, 
though the curves are shifted with respect to each other due to the different behavior at intermediate times.\\
More pronounced differences are seen in the local roughness $G(1,t)$, plotted in Fig.~\ref{fig:5roughness}(b).
This is expected, since $G(1,t)$ is more sensitive to the differences in surface morphology seen in Figs.~\ref{fig:1trajectories} and \ref{fig:4hh_corr}. 
One main feature of the C$_{60}$ growth is the rounded island structure. 
This leads to a slower increase of the local roughness compared to the Ag system, where the islands are dendritic. On the other hand, the Ag system is characterized by a layer-by-layer-like growth at intermediate times. This is reflected by the longer plateau on the corresponding $G(1,t)$ in the corresponding range of times. However, 
for long times the local 
roughness of C$_{60}$ grows faster than for Ag, resulting 
in very similar values of $G(1,t)$ at long times. This can be interpreted as an indication for similar morphologies in the two systems during the late stages of growth.

\subsection{Local Dynamics}\label{Sec:3.2Local}
\begin{figure}%

      \includegraphics[scale=0.45]{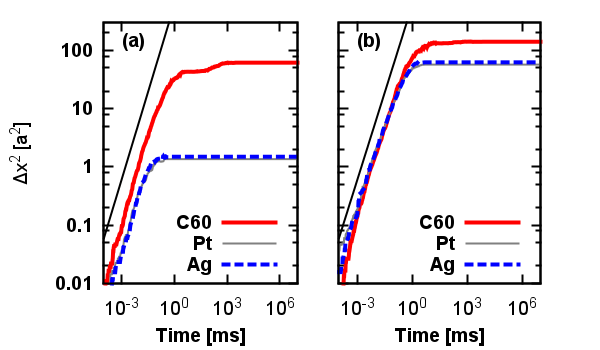}

\caption{(color online) The MSD of particles that arrive (a) between islands or (b) on top of islands after the growth of 0.5\,ML for different interaction energies $E_{\textrm{n}}(\textrm{C}_{60})=0.13$\,eV, $E_{\textrm{n}}(\textrm{Ag})^{,}=0.72$\,eV and $E_{\textrm{n}}(\textrm{Pt})^{,}=0.92$\,eV. 
All systems are simulated on a triangular lattice with equal energy barriers $E_{\textrm{free}}=0.54$\,eV and $E_{\textrm{ES}}=0.11$\,eV. The turquoise lines represent the MSD for free diffusion (linear time dependence).}%
\label{fig:2MSD}%
\end{figure}

\begin{center}
\begin{figure*}%
\begin{minipage}{6.5cm}%
      \includegraphics[scale=0.5]{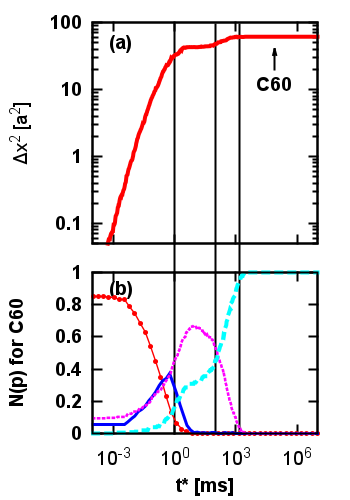}
\end{minipage}
\begin{minipage}{8cm}%
      \includegraphics[scale=0.5]{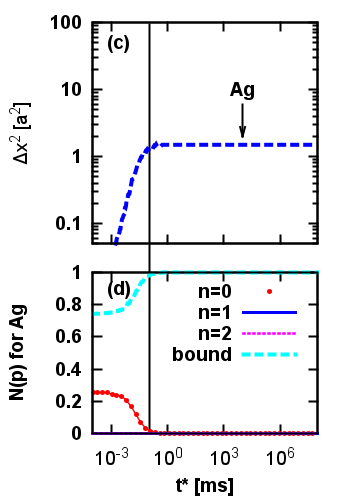}
\end{minipage}%
\caption{(color online) Part (a) and (c) show the MSD of the C$_{60}$ and Ag systems already plotted in Fig.~\ref{fig:2MSD} (a) (i.e. diffusion in between islands), while (b) and (d) correlate these data to the time dependence of the fraction 
of particles doing a certain process. These processes can be free diffusion ($n=0$) and diffusing away from a site with $n=1\textrm{ or }2$ 
neighbor particles. Also plotted are the fractions of particles that are immobilized (bound). All other processes are negligible.}%
\label{fig:3processe}%
\end{figure*}
\end{center}

We now turn to the dynamics on the particle-resolved level. Figure~\ref{fig:2MSD} depicts the MSD after the 
growth of 0.5\,ML, where we distinguish between particles that arrive between [Fig.~\ref{fig:2MSD}(a)] and on [Fig.~\ref{fig:2MSD}(b)] islands. 
All curves share the same general structure in that the MSD initially grows in time (with exponent $\approx 1$) and then saturates, indicating immobilization. 
However, when comparing the curves for arrivals between and on islands, the two systems behave differently. Indeed, the diffusion behavior of the Ag system is quite sensitive to the 
location of a particle's arrival, whereas that of C$_{60}$ is not.
We now relate these features to the morphology.\\
In the atomic system, the islands are fairly dendritic [see Fig.~\ref{fig:1trajectories}(b)]. As a consequence, particles that arrive between the islands [Fig.~\ref{fig:2MSD}(a)] 
travel only small distances before they encounter
the edge of an island. In this situation the majority of atomic particles either attaches for long time spans or becomes immobilized completely, as further particles attach before they can detach themselves.
This leads to an early onset of sub-diffusive behavior (for $t^{\ast}>0.05$\,ms) and average travel distances of just a few nanometres before the particles are immobilized (see plateau in the MSD).\\
A quite different behavior is seen for atomic particles that arrive on the islands [see Fig.~\ref{fig:2MSD}(b)]. These particles can diffuse across fairly large dendritic structures before encountering other particles. 
The step
edges hinder particles from leaving the island, but they do not noticably slow down their motion on the islands. 
Therefore, atomic particles on the islands can 
travel significantly further (as compared to the case discussed before) before they become immobilized and the MSD saturates.\\
The C$_{60}$ system is characterized by a completely different island morphology [see Fig.~\ref{fig:1trajectories}(a)]. As a consequence, molecules that arrive between islands can diffuse 
far further before encountering other molecules, just because there is more free space. Moreover, C$_{60}$ molecules can \textit{detach} after encountering other molecules (as a result of the weaker binding).
These effects lead to much larger traveled distances both between and on islands. Furthermore, the MSD curves for these two cases are similar. \\
An even better understanding of the systems emerges when we relate the MSDs to the occurrence of certain individual processes. This is done in Fig.~\ref{fig:3processe}, where we focus on particles arriving in between islands. 
Parts (a) and (c) show the corresponding MSDs, while parts (b) and (d) contain data for the fraction of particles involved in a specific process, $N(p,t)$
For both, C$_{60}$ and Ag systems it is clearly visible that the early stages of growth are dominated by freely diffusing particles ($n=$0).
These form the main contribution to the MSD at small times.
However, once particles begin to interact, distinct differences between C$_{60}$ and Ag become apparent.
During the growth of C$_{60}$ there are many events where particles detach from one or two neighboring  
particles (see curves in Fig.~\ref{fig:3processe}(b) with $n=1,2$). Such events are absent in the Ag system. We understand this difference as a consequence of the larger binding energy in the Ag system.

\section{Conclusion}\label{sec:conclusion}

In this paper we have discussed the single-particle and global dynamics accompanying the surface growth of the nanomolecular system C$_{60}$. One main goal in this context was to understand 
the similarities and differences between nanomolecular and atomic growth for the spherical molecule C$_{60}$. 
To this end, we have first identified energetic differences between the two system types. We then used the two system types to study both 
single-particle trajectories and the overall surface morphology.\\
We have found that there are indeed pronounced differences in the surface morphology during growth: The C$_{60}$ system displays compact islands with 
a rather large fractal dimension and 
significant spatial correlations between the islands. In contrast, the fractal dimension and range of ordering are far smaller in atomic systems. These differences in the 
global surface morphology during 
growth can be traced back to the differences in the energy barriers that single particles have to overcome. Moreover, we have shown that the differences in morphology is intimately 
related to the single-particle dynamics. Atomic particles diffusing between islands can cover only small distances before they are immobilized, because the islands are fairly 
dendritic. On the other hand, particles in the C$_{60}$ system can diffuse quite far, since the compact islands are separated
through large, free surfaces. \\
The large diffusion distances are effectively enhanced by the fact that due to the small binding energies, C$_{60}$ 
molecules can detach from island edges before they are immobilized. However, unlike colloidal systems \cite{Ganapathy2011}, C$_{60}$ has to overcome an energetic atom-like step-edge barrier for interlayer diffusion.
This leads to a reduced mobility of both, molecules and atoms, between layers. The reduced mobility then leads to a roughening of the surface on large time-scales.\\
In summary we find a complex interplay of single-particle and global dynamics, whose characteristics reflect special molecular features of C$_{60}$, in particular the relatively small effective binding energy. 
Starting from our findings, one interesting question for further studies would be the connection between single-particle-trajectories and the long-time scaling behavior derived from continuous
rate-model descriptions. \\
Further, for a more detailed insight concerning the interplay of energy landscape and dynamics, it may be interesting to couple KMC simulations and MD
simulations. Withing such a study, configurations gained 
from KMC could be used  to generate inital configurations for MD simulations. Conversely, MD simulations could be used calculate ``on the fly'' the energy parameters required in the KMC simulations,
thus taking into account the time-dependence of the energetics. Such an approach would yield important insight into all time-scales 
from up to $\mathcal{O}(10^{-9})$s towards $\mathcal{O}(10^4)$s.\\
Experimentall, real-time measurements of the overall surface morphology during the growth of organic molecules are possible through x-ray scattering \cite{Unseres} 
or low energy electron microscopy \cite{Heringdorf2001,Khokhar2012}. In contrast, the dynamics of individual molecules is experimentally not (yet) accessible, as these 
phenomena take place on very small time-scales. Very recently, measurements of the particle-resolved dynamics of colloidal particles at room temperature have been performed \cite{Ganapathy2011}. 
However, in molecular systems, particle resolved dynamics under consideration of the local particle neighborhood is only accessible at very low temperatures. 
An example are scanning tunneling microscopy experiments to track the motion of individual molecules on a substrate and to determine waiting times \cite{Kumagai2008}. Extending such investigations to higher temperatures
may open the path to quantities such as the ones calculated in the present work.

\section{Acknowledgements}\label{Sec:5Acknows}
We gratefully acknowledge stimulating discussions with S.~Bommel and S.~Kowarik. 
This work was supported by the Deutsche Forschungsgemeinschaft within the framework
of the Collaborative Research Center CRC 951 (project A7).

\appendix
\section{Comparing C$_{\textrm{60}}$ energy parameters to literature}\label{app1}

The energy parameters listed in Table~\ref{tab:1values}, which we obtained by comparing with corresponding experiments \cite{Unseres}, are of the same order of magnitude as other values
\cite{Korner2011, Liu1990} reported in the literature, but differ in their actual magnitude. In the following we briefly discuss to which end these differences
can be attributed to differences of energy barrier definitions and of simulation approaches.\\
We start by considering the energy barrier stemming from nearest-neighbor interactions. 
The corresponding value quoted in the KMC study of K\"orner \textit{et al. }\cite{Korner2011} is $E_{\textrm{b}}^{\textrm{K\"orner}}=0.271$\,eV. This equals the depth of the pair interaction
potential of two interacting C$_{\textrm{60}}$ molecules, as derived by Girifalco \cite{doi:10.1021/j100167a002}. However, particles in the simulation of K\"orner \textit{et al. }need 
to overcome $nE_{\textrm{b}}/2$ to move from a site with $n$ neighboring particles to a site with no neighboring particles. Therefore the definition of $E_{\textrm{b}}$ differs
from our definition of $E_{\textrm{n}}$. To be correct, we have to we compare $E_{\textrm{b}}^{\textrm{K\"orner}}/2=0.1355$\,eV with our value $E_{\textrm{n}}=0.13 \pm 0.02 $\,eV.
Clearly, these are in very good agreement.\\
Next, we consider the free diffusion energy. 
Both K\"orner \textit{et al. }\cite{Korner2011} and Liu \textit{et al.} \cite{Liu1990} report a value $E_{\textrm{free, K\"orner}}=0.178$\,eV; however, they also 
use an attempt frequency of $\nu=2\cdot10^{11}$\,Hz. Moreover, both studies are based on a hexagonal lattice under consideration of interstitial sites. Here, we neglect these sites, yielding
a somewhat coarse-grained approach. We note that without the coarse-grained approach it would not be possible to simulate such a large system for minutes 
to hours of experimental time. In one diffusion step on our coarse-grained lattice a particle overcomes two times the barrier $E_{\textrm{free, K\"orner}}=0.178$\,eV. 
In addition there three options to diffuse from the interstitial site. Since only one option leads to our coarse-grained destination site,
an additional geometric factor of $1/3$ needs to be included in the diffusion rate. \\
Taking, furthermore, the difference in the attempt frequency $\nu$ into account, 
we gain the following estimate of a coarse-grained free diffusion barrier from the values reported in \cite{Korner2011, Liu1990}
\begin{align}
 \notag E_{\textrm{free, K\"orner}}^,&\approx -\textrm{ln}\left(\frac{1.4\cdot 10^{13} \textrm{ Hz}}{2 \cdot 10^{11} \textrm{\,Hz}}\right) kT - \textrm{ln}\left(\frac{1}{3}\right) k T\\
 \notag&\qquad+ 2\cdot E_{\textrm{free, K\"orner}}\\
 \notag&\approx 0.122 \textrm{\,eV}+0.032 \textrm{eV}+0.356 \textrm{\,eV}\\
      &\approx 0.51 \textrm{\,eV} \textrm{,}
\end{align}
which lies within the error margins of our value $E_{\textrm{free}}$ (see Table~\ref{tab:1values}). This estimate was gained using $T=60^{\circ}$C.\\
Finally, our value of the Ehrlich-Schwoebel barrier $E_{\textrm{ES}}=0.11\pm0.02$\,eV (see Table~\ref{tab:1values}) is in very good agreement with values derived from density-functional
theory calculations for step edge barriers  $E_{\textrm{ES}}^{Goose}\approx0.104$\,eV \cite{Goose2010}.

\section{Comparison between C$_{\textrm{60}}$ and atomic systems}\label{app2}

The comparison to atomic systems considered in this work is made possible through the work of other groups, in which the energy parameters listed in table \ref{tab:1values}
were successfully employed to simulate atomic growth on a coarse-grained lattice such ours. 

\subsubsection*{Pt on Pt(111)}
Hohage \textit{et al.} \cite{Hohage1996} used a free diffusion energy of $E_{\textrm{free,Pt}}=0.26$\,eV and a attempt frequency of $\nu=5\cdot10^{12}$\,Hz to 
simulate the growth of Pt on Pt(111). They employed a simulation grid that only contains sites that are occupied in a bulk crystal.
This approach to lattice coarse-graining is equivalent to ours, which enables a comparison between diffusion energies. Specifically, we compare the energy 
$E_{\textrm{free,Pt}}^{,}$ to the free diffusion energy $E_{\textrm{free}}$ of 
C$_{60}$, where $E_{\textrm{free,Pt}}^{,}$ is related to $E_{\textrm{free,Pt}}$ \cite{Hohage1996} via the attempt frequency 
\begin{equation}
 E_{\textrm{free,Pt}}^{,} =0.26\textrm{eV}-\textrm{ln}\left(\frac{5\cdot 10^{12}\textrm{Hz}}{1.4 \cdot 10^{13}\textrm{\,Hz}}\right) kT\approx0.29\textrm{\,eV.}
\end{equation}
Comparing this value to neighbor interaction energies mentioned in the study by Feibelman and Michely \cite{Feibelman2001L723}, who found $E_{b,2}=0.5$\,eV, we obtain a ratio  
\begin{equation}
 R(\textrm{Pt})=\frac{E_{\textrm{free,Pt}}^,}{E_{\textrm{free,Pt}}^,+E_{b,2}}\approx 0.37 \textrm{.} \label{eqn::R_Pt}
\end{equation}

\subsubsection*{Ag on Ag(111)}
The values for free diffusion energy barriers for Ag/Ag(111) reported in the literature show a wider spread, presumably due to the tendency of Ag to oxidize and the influence of this impurity 
on measurements. Values quoted range from $E_{\textrm{free,Ag}}=0.1$\,eV with $\nu=10^{11}$\,Hz \cite{Li2009} and $E_{\textrm{free,Ag}}=0.097$\,eV with $\nu=2\cdot 10^{11}$\,Hz \cite{Brune1995}, via the combination of $E_{\textrm{free,Ag}}\approx0.067$\,eV with $\nu=10^{12}$\,Hz \cite{Nandipati2009, Latz2012}
to $E_{\textrm{free,Ag}}=0.1$\,eV with $\nu=10^{13}$\,Hz \cite{Li2008} and the combination of $E_{\textrm{free,Ag}}=0.12$\,eV with $\nu=10^{13}$\,Hz \cite{Blackwell2012}. All of the quoted values have been used to 
study Ag on Ag(111) using kinetic Monte-Carlo simulations on a coarse-grained lattice. \\
In view of this spread, we have considered an intermediate value for the diffusion energy-barrier, which was determined for pure Ag using Molecular Dynamics and nudged-elastic band approaches:
\begin{equation}
 E_{\textrm{free,Ag}}^{,} =0.067\textrm{eV}-\textrm{ln}\left(\frac{10^{12}\textrm{\,Hz}}{1.4 \cdot 10^{13}\textrm{Hz}}\right) kT\approx0.143\textrm{\,eV.}
\end{equation}
Similarly the range of neighbor interaction energies ranges from $E_{\textrm{n}}=0.15$\,eV to $E_{\textrm{n}}=0.24$\,eV (\cite{Evans20061} and references within), while most studies appear to agree on $E_{\textrm{n}}\approx 0.19$\,eV.
Using these values we find:
\begin{equation}
 R(\textrm{Ag})=\frac{E_{\textrm{free,Ag}}^,}{E_{\textrm{free,Ag}}^,+E_{\textrm{n}}}\approx 0.43 \textrm{.} \label{eqn::R_Ag}
\end{equation}


\end{document}